# Review of Best Practice Methods for Determining an Electrode Material's Performance for Ultracapacitors

Meryl D. Stoller and Rodney S. Ruoff*

*Department of Mechanical Engineering and the Texas Materials Institute, The University of Texas at Austin, One University Station C2200, Austin, Texas, 78712-0292 USA*
* Corresponding Author. Email: r.ruoff@mail.utexas.edu

**Abstract**

Ultracapacitors are rapidly being adopted for use for a wide range of electrical energy storage applications. While ultracapacitors are able to deliver high rates of charge and discharge, they are limited in the amount of energy stored. The capacity of ultracapacitors is largely determined by the electrode material and as a result, research to improve the performance of electrode materials has dramatically increased. While test methods for packaged ultracapacitors are well developed, it is often not feasible for the materials scientist to assemble full sized, packaged cells to test electrode materials. Methodology to reliably measure a material's performance for ultracapacitor electrode use is not well standardized with the different techniques currently being used yielding widely varying results. In this manuscript, we review the best practice test methods that accurately predict a materials performance, yet are flexible and quick enough to accommodate a wide range of material sample types and amounts.

## Introduction

Ultracapacitors based on electrochemical double layer capacitance (EDLC) are electrical energy storage devices that store and release energy by nanoscopic charge separation at the electrochemical interface between an electrode and an electrolyte[1]. While the charge storage mechanism of EDLCs is based on the interfacial double-layer of high specific area carbons, another class of capacitors is based on pseudocapacitance, and thus associated with electrosorption and surface redox processes at high surface area electrode materials such as metal oxides and conducting polymers. Hybrid capacitors are the combination of a faradic battery-type electrode coupled with a capacitive electrode in a two electrode module (termed an asymmetric capacitor)[2]. While the energy density of ultracapacitors is very high compared to electrostatic and electrolytic capacitors, it is still significantly lower than batteries and fuel cells. Coupling with batteries (or another power source) is still required for supplying energy for longer periods of time. Thus, there is a strong interest as enunciated, e.g., by the U.S. Department of Energy, for increasing the energy density of ultracapacitors to be closer to the energy density of batteries[3].

The electrode material is a key component that determines an ultracapacitor's capacity and the most definitive test for a new electrode material is how it performs in a full scale, commercial ultracapacitor. However, it is not always practical to use a full sized, packaged cell, especially when dealing with minute quantities of material and/or involving a large number of different types of samples to be tested. The goal of this manuscript is to review experimental procedures that accurately evaluate a material's performance, yet are flexible and rapid enough to accommodate a large number of samples over a wide range of material types and quantities. Test results should also be repeatable and match those from other locations and research groups. At this time, the measurement methods for determining a material's performance are not well standardized and as a result, it is difficult to assess the true performance reported in the literature, which in our opinion is hindering progress in this field.

Methodology for electrode material testing can be grouped into *test fixture configuration* and *measurement procedures*. Test fixture configuration includes the **test fixture type** along with guidelines for **electrode mass** and **thickness**, and other cell components including the **electrolyte**, **separator**, **current collectors**, and **binder**. Measurement procedures include **electrochemical measurements and parameters** along with the **computations to reduce the data to the desired metrics**.

## Test Fixture Configuration

A typical ultracapacitor unit cell is comprised of two electrodes that are isolated from electrical contact by a porous separator[1]. Electrodes often contain conductive but low surface area additives such as carbon black to improve electrical conductivity. Current collectors of metal foil or carbon filled polymers are used to conduct electrical current from each electrode.  The separator and the electrodes are impregnated with an electrolyte, which allows ionic current to flow between the electrodes while preventing electronic current from discharging the cell.  A packaged ultracapacitor module, depending upon the desired size and voltage, is constructed of multiple repeating unit cells.

A test fixture configuration that closely mimics the unit cell configuration will more closely match the performance of a packaged cell.  Two-electrode test fixtures are either available commercially or can be easily fabricated from two stainless steel plates as shown in Figure 1[4].  Three-electrode electrochemical cells are commonly used in electrochemical research and consist of a working electrode, a reference electrode, and a counter electrode. *Three-electrode cells differ from two-electrode test and packaged cells in several important aspects*.  With the three-electrode configuration, only one electrode, called the working electrode, contains the material being analyzed and the applied voltage and charge transfer across the single electrode is markedly different than with a two-electrode cell configuration.  The potential drop across the single electrode/electrolyte interface is controlled with respect to a reference electrode, however the current flow is from the working electrode to the counter electrode. Khomenko, et al reported the dependence of measured capacitance values on test cell configuration[5]. Composite electrodes comprised of MWCNT's and conducting polymers were measured using both two-electrode and three-electrode cell configurations.  Figure 2 shows the cyclic voltammograms (CV) for three- and two-electrode configurations using such PANI/MWNT electrodes.  In the case of three-electrode cell measurements, values of 250 to 1100 F/g were measured.  For the same materials in a two-electrode cell, values of 190 to 360 F/g were measured.  Table 1 lists the specific capacitance results for two different materials (PPy and PANI) as measured with each cell type.  As seen from the table, the three-electrode cell yields values approximately double those of the two electrode cell.  While valuable for analyzing the faradic reactions and voltages at a single surface, *the heightened sensitivity of the three-electrode configuration can lead to large errors when projecting the energy storage capability of an electrode material for ultracapacitor use*.

The mass of the active material and thickness of the electrodes also influences the measured results. Depending upon whether an ultracapacitor is constructed to optimize energy density or power density, commercial cell electrode thicknesses range from about 10 um thick (high power density) to several hundred microns thick (high energy density). Test electrodes should be of comparable thicknesses *and ones that are extremely thin and/or contain very minute amounts of material can lead to an overstatement of a material's performance*. Signal to noise is also a concern. The capacitance of the material in the electrodes should be significantly higher than that contributed by other cell components such as the electrode support surface, collector, and other conducting surfaces within the test fixture. In addition, mass measurement errors can be significant when handling and weighing micro gram sized electrodes. For reliable measurements, a test cell should have a capacitance of 0.25 or more Farads with the mass of the active material on the order of 10 or more milligrams. Hu et al. demonstrated the dependence of mass and thickness on measured results[6]. Figure 3 shows specific capacitances for four different mass loadings for electrodes constructed of single-walled carbon nanotube (SWCNT) paper and measured in a two-electrode test cell[6]. The graph shows that, with aqueous electrolyte, as the mass loading was increased from 72 μg/cm$^2$ to 1.33 mg/cm$^2$, the specific capacitance at a reported scan rate of about 2 A/g decreased from about 200 to 85 F/g respectively. At a scan rate of 5 A/g and for the same two masses, the specific capacitance decreased from about 175 to 75 F/g. The reported electrode thickness for the 1.33 mg/cm$^2$ mass loading was 14 μm[7]. The electrodes with a mass loading of 72 μg/cm$^2$ thus have a thickness of about 0.75 micron, an order of magnitude thinner than commercial ultracapacitor electrodes. It is important to use appropriate electrode thicknesses and masses for any meaning to be attached to reported values of specific capacitance and energy density.

The most common organic electrolytes are tetraelthylammonium tetrafluoroborate (TEABF4) in either propylene carbonate (PC) or acetonitrile (AN). Common aqueous electrolytes include KOH and $H_2SO_4$. Since energy stored is related to the square of voltage, organic electrolytes are currently used in commercial ultracapacitors due to their wider electrochemical window (about 2.7 volts) as compared to about 1 volt for aqueous electrolytes. Ionic liquid electrolytes are also being adopted due to their increased electrochemical windows and improved thermal stability. A material's performance with an aqueous electrolyte will typically yield higher specific capacitances and does not indicate its performance with an organic or IL electrolyte. Figure 3 shows SWCNT paper electrodes of equal mass measured with aqueous (paper:$H_2SO_4$:1V:72μg/cm$^2$) and organic (paper:organic:1V:72 ug/cm$^2$) electrolytes[6]. The values for aqueous electrolytes are consistently 40 - 50% higher than with the organic electrolyte over a wide range of scan rates. The performance disparity for different electrolytes also depends upon material type and morphology. Table 2 shows specific capacitances for electrodes composed of chemically modified graphene material[4]. For this material, the specific capacitance differences due to the electrolyte have in our work ranged from about 20 to 25 percent higher for the aqueous electrolyte. Other cell components such as binders, current collectors, and separators also have an affect upon cell performance. However, when from a commercial source, their impact upon measured values is relatively small.

**Measurement procedures**

Charging rates, voltage ranges, and methods for calculation of metrics also affect the reported results and, when possible, should match currently established and accepted procedures used for packaged cells. The primary performance metrics for packaged ultracapacitors include gravimetric energy and power densities, and life cycle testing[8-11]. In turn, an ultracapacitor's energy density (W-hr/kg) is primarily determined by the cell's electrode material and electrochemical voltage window. With energy density currently the primary limitation for ultracapacitors, the most important metric for an electrode material is thus its specific capacitance (F/g). An ultracapacitor's power scales with the square of voltage divided by its equivalent series resistance (ESR)[12]. The measured ESR of a test cell (as well as that of a full scale packaged capacitor) is due to all cell components (leads, current collectors, electrodes, electrolyte, separator) and therefore only a portion of the measured resistance can be attributed to the electrode material itself. Other metrics such as an electrode material's energy and power density also do not correlate directly to that of a packaged cell and must include information such as package dimensions and the mass of the other cell components to be meaningful.

Specific capacitance is the capacitance per unit mass for one electrode (equation 1)

$$C_{sp} \text{ (F/g)} = 4 * C / m \quad (1)$$

where $C$ is the measured capacitance for the two-electrode cell and $m$ is the total mass of the active material in both electrodes. The multiplier of 4 adjusts the capacitance of the cell and the combined mass of two electrodes to the capacitance and mass of a single electrode. If volume is more important for the targeted application, the electrode material's volume may be substituted for mass. Cell capacitance is best determined from galvanostatic or constant current (CC) discharge curves using equation 2 with I the discharge current and

$$C = I / (dV/dt) \quad (2)$$

$dV/dt$ calculated from the slope of the CC discharge curve. The same voltage range should be used for testing as that used for commercial cells and should reflect the electrolyte's electrochemical window - from 0 V to approximately 1 V for aqueous and from 0 V to 2.5-2.7 V for organic electrolytes. Maximum voltages for hybrid cells will depend upon electrode materials and electrolytes. The initial portion of a discharge curve exhibits an *IR* drop due to internal resistance and the rest of the curve will typically be linear for non-faradic materials. Pseudocapacitive and hybrid systems can exhibit large deviations in linearity based upon varying capacitance with voltage. Figure 4 shows CC charge-discharge curves (100 mA/g) of an asymmetric manganese oxide/activated carbon ultracapacitor in 2 mol/L $KNO_3$ electrolyte cycled at different maximum cell voltages[13]. When the maximum voltage is at 2.2 V, the CC curve is no longer symmetric indicating non-capacitive behavior. Figure 5 shows, for the same cell, the coulombic efficiency and specific capacitance (F/g) vs. maximum voltage[13]. While the specific capacitance continues to increase with increasing voltage range, the coulombic efficiency decreases dramatically when cycled above 2 V. Driving a cell above its true maximum operating voltage can lead to an overestimation of specific capacitance and cells operated at these levels will have shortened lifetimes and poor efficiencies due to the non-reversible reactions within the cell. Significant errors can also be introduced by the method used to calculate the slope ($dV/dt$). As stated previously, capacitance varies with voltage, especially for hybrid and pseudocapacitive cells, and it is important to calculate capacitance using the typical operating voltage range for the application that the device will be used. Most ultracapacitors will be operated in the range of $V_{max}$ to approximately ½ $V_{max}$ and the recommended method is to use two data points from the discharge curve with $dV/dt = (V_{max} - ½ V_{max}) / (T_2 - T_1)$. *Including the lower half of the voltage range in the calculations can distort the apparent capacitance above that which is practically realizable for an actual application.*

*Very low rates of discharge also lead to large errors, especially when coupled with small electrode masses,* with cell leakage or capacitance from other components significantly distorting the actual capacitance contributions from the electrode material. Charge and discharge rates should be specified in units of current per electrode mass with the duration of charge and discharge corresponding to typical ultracapacitor applications. Current should be adjusted to provide charge and discharge times of approximately 5 to 60 seconds[12]. For example, a test cell with two 10 mg electrodes composed of 100 F/g specific capacitance material will have a capacitance of 0.5 F. With a discharge current of 40 mA, corresponding to a discharge density of 4 A/g, discharge time from 2.7 to 0 volts will be approximately 34 seconds. Figure 3 shows the dependence of specific capacitance on the rate of discharge and electrode mass loading[6]. For the electrode labeled "paper:$H_2SO_4$, 1.33 mg/cm$^2$" with reported electrode thickness of 14 μm[7], the measured specific capacitance varies significantly (from over 120 F/g to about 85 F/g) for discharge rates of less than 2 A/g. This effect is most pronounced with the thicker electrode highlighting the importance of using electrode thicknesses that match packaged cells.

While CC data is recommended, CVs may be used to calculate capacitance. Using equation 2 and CV data, *I* is the average current during discharge (from Vmax to zero volts) and *dV/dt* is the scan rate. As with CC curves, capacitance depends on scan rate, voltage range, and computation method. The cell should be cycled for 20 or more cycles prior to recording the data and should only be cycled from 0 volts to the maximum voltage. Figure 6 shows two CVs, one cycled from 0 V to 1 V (top), the second is the same cell cycled from -1 V to 1 V (bottom). The first and the 20$^{th}$ cycles are shown on each CV. When a cell is first cycled or when it is cycled from a negative to positive voltages, there are increased current levels due to reversing the polarity of the cell.

Table 4 shows the results of various calculation methods for different ranges of *I* from the two CV curves with over a 300 percent difference in values of specific capacitance. *As with CC curves, the discharge rate should reflect that of a typical ultracapacitor application. Voltage scan rates of at least 20 to 40 mV/sec are needed to maintain discharge times on the order of a minute and adequately reflect a material's performance.*

**Secondary Material Performance Metrics**

A packaged cell's specific power and cycle life depend upon all components within the cell as well as the cell architecture. The main indicator for the power capability for a packaged cell is based upon its direct current resistance or ESR. For packaged cells, ESR is typically determined from the CC tests using the *IR* drop at the beginning of the discharge curve and the same method should be used for test cells. When the discharge current (I) is initiated at the beginning of the discharge cycle, the ESR is the value calculated from the change in voltage (*IR* drop) divided by the step in current. The reader is referred to Burke[12] and Zhao[11] for a detailed description. ESR depends upon cell size and the capacitance of the test cell should also be reported. Cell life also depends upon all cell components and a simple constant power or constant current cycle is adequate for an initial gauge of a material's stability. One should note that since degradation of a material-electrolyte system occurs primarily at higher voltage, any life cycle testing should include the maximum rated voltage in each cycle. Again the reader is referred to Burke[12] for more detailed methods of testing packaged cells.

**Summary of recommendations**

While a three-electrode cell is valuable for determining electrochemical-specific material characteristics, a two-electrode test cell mimics the physical configuration, internal voltages and charge transfer that occurs in a packaged ultracapacitor and thus provides the best indication of an electrode material's performance. For good signal to noise and to minimize measurement errors, the test cell should have a capacity of over 0.25 farad with the mass of the active material on the order of 10 or more mg. Electrode thicknesses should be on the order of packaged commercial cells (>15 μm thick) and should be tested with the same electrolytes that would be used in an actual capacitor.

Charge and discharge rates (> 2 A/g and > 20 mV/sec) and operating voltages (>2.5 V for organic electrolytes) should match that of typical ultracapacitor applications. Calculations for capacitance and ESR should match those currently recommended for commercial ultracapacitors.

Any reporting of energy and power densities should thus be done in the context of a full sized, packaged ultracapacitor.  As electrode thickness affects measured performance, ESR measurements should only be performed using test electrodes with the same thicknesses of commercial cells.  Energy and power density calculations should include the mass of all components including the package.  Lifetime testing should be performed using a cycle that includes the maximum rated voltage.

**Conclusion**

Measurement methods for determining a material's performance for use as an ultracapacitor electrode are not well standardized with different techniques currently being employed leading to wide variations in reported results.  The various experimental procedures were reviewed and best practice methods were recommended that effectively simulate a packaged, commercial cell, yet are flexible and rapid enough to accommodate a large number of samples with a wide range of material types and quantities.  We believe adoption of these measurement practices will enable the more accurate determination and reporting of an electrode material's performance.

**Acknowledgement**

This work was supported by the U.S. DoE SISGR under Award DE-ER46657, by Graphene Energy, Inc., and by a startup package to RSR by The University of Texas at Austin.

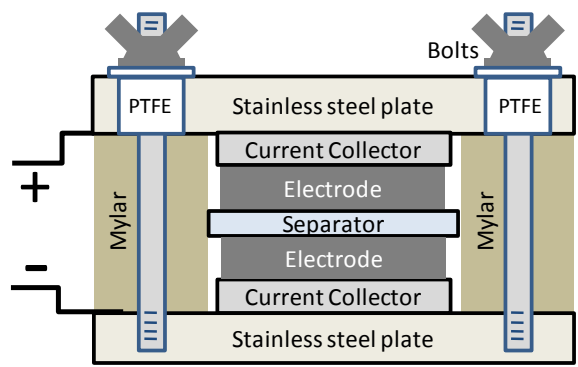

Figure 1. Two electrode test cell configuration.[4]

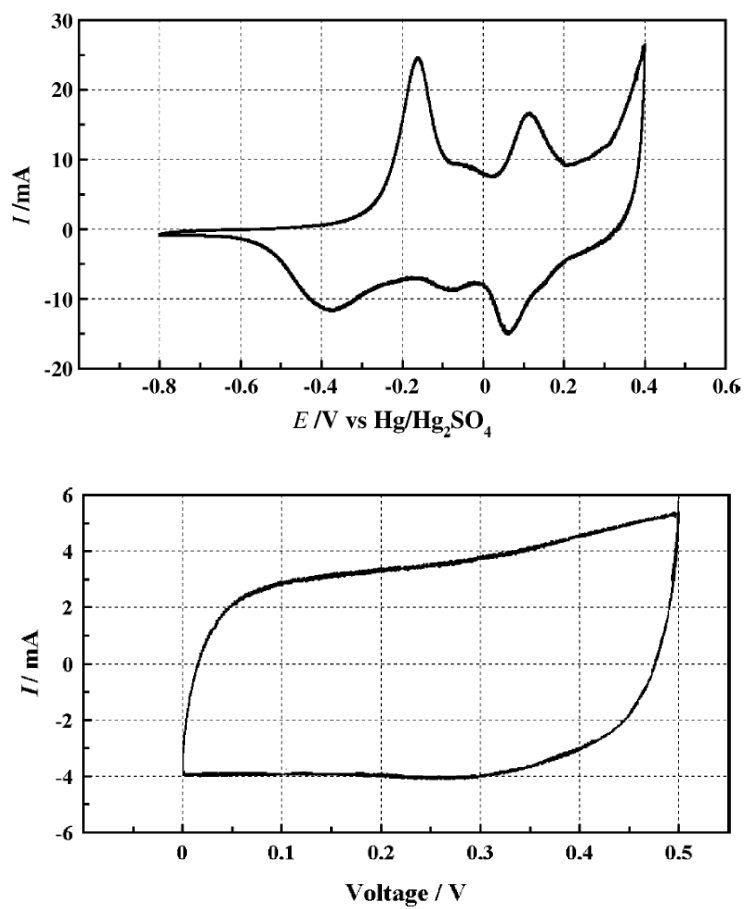

Figure 2. (top) CV of PANI/MWNTs electrode using a three electrode cell and (bottom) using a two electrode cell. Reprinted from V. Khomenko, *Electrochim Acta* **50**, 2499-2506 (2005) with permission from Elsevier.[5]

Table1. Values of specific capacitance (F/g) depending on cell type. Reprinted from V. Khomenko, *Electrochim Acta* **50**, 2499-2506 (2005) with permission from Elsevier.[5]

| ECP in the composite electrode | Three-electrode cell | | Two-electrode cell | |
|---|---|---|---|---|
| | CV | Galvanostatic discharge | CV | Galvanostatic discharge |
| PPy | 506 | 495 | 192 | 200 |
| PANI | 670 | 650 | 344 | 360 |

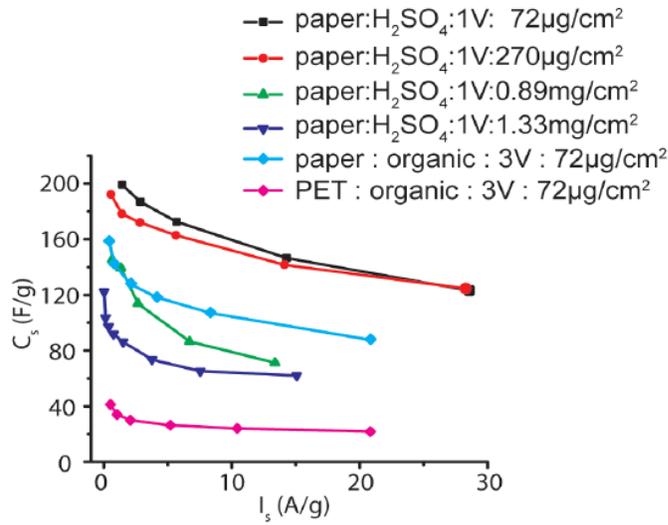

Figure 3. Measured capacitance decrease with an increase of electrode mass. Reprinted from L.B. Hu, *P Natl Acad Sci USA* **106**, 21490-21494 (2009).[6]

Table 2. Graphene-based electrode performance by electrolyte and measurement method.[4]

| Electrolyte | Galvanostatic discharge | | Cyclic Voltammogram average | |
|---|---|---|---|---|
| | mA | | mV/sec | |
| | 10 | 20 | 20 | 40 |
| KOH | 135 | 128 | 100 | 107 |
| TEABF$_4$/PC | 94 | 91 | 82 | 80 |
| TEABF$_4$/AN | 99 | 95 | 99 | 85 |

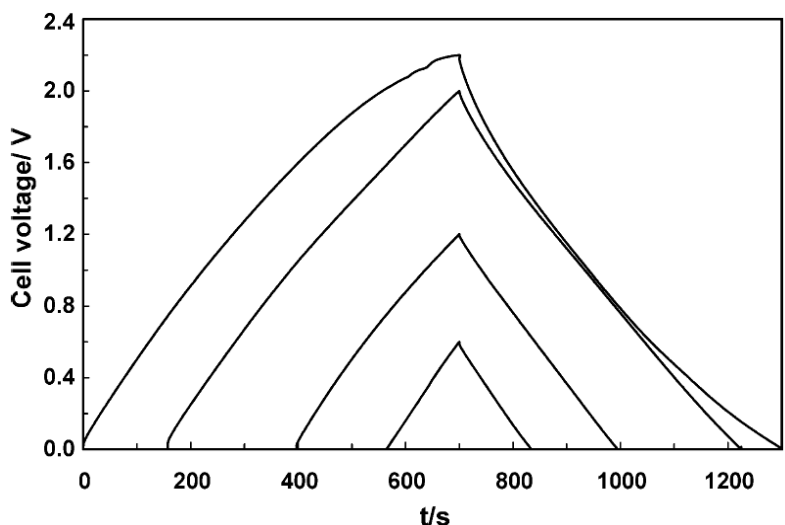

Figure 4. CC charge-discharge curves (100 mA/g) of an asymmetric manganese oxide/activated carbon ultracapacitor in 2 mol/L $KNO_3$ electrolyte. Reprinted from V. Khomenko, *J Power Sources* **153**, 183-190 (2006) with permission from Elsevier.[13]

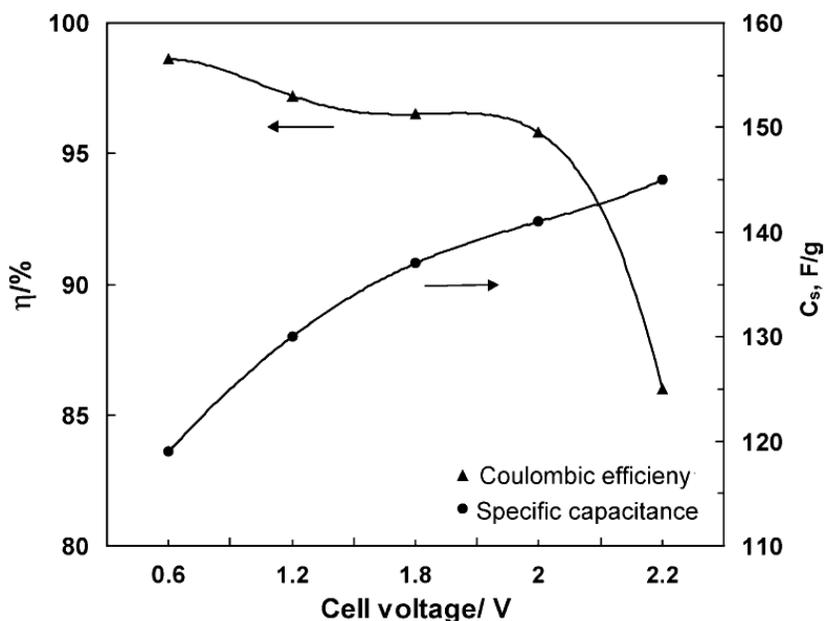

Figure 5. Coulombic efficiency and specific capacitance (F/g) of an asymmetric manganese oxide/activated carbon ultracapacitor in 2 mol/L $KNO_3$ electrolyte vs. the cell voltage. Reprinted from V. Khomenko, *J Power Sources* **153**, 183-190 (2006) with permission from Elsevier.[13]

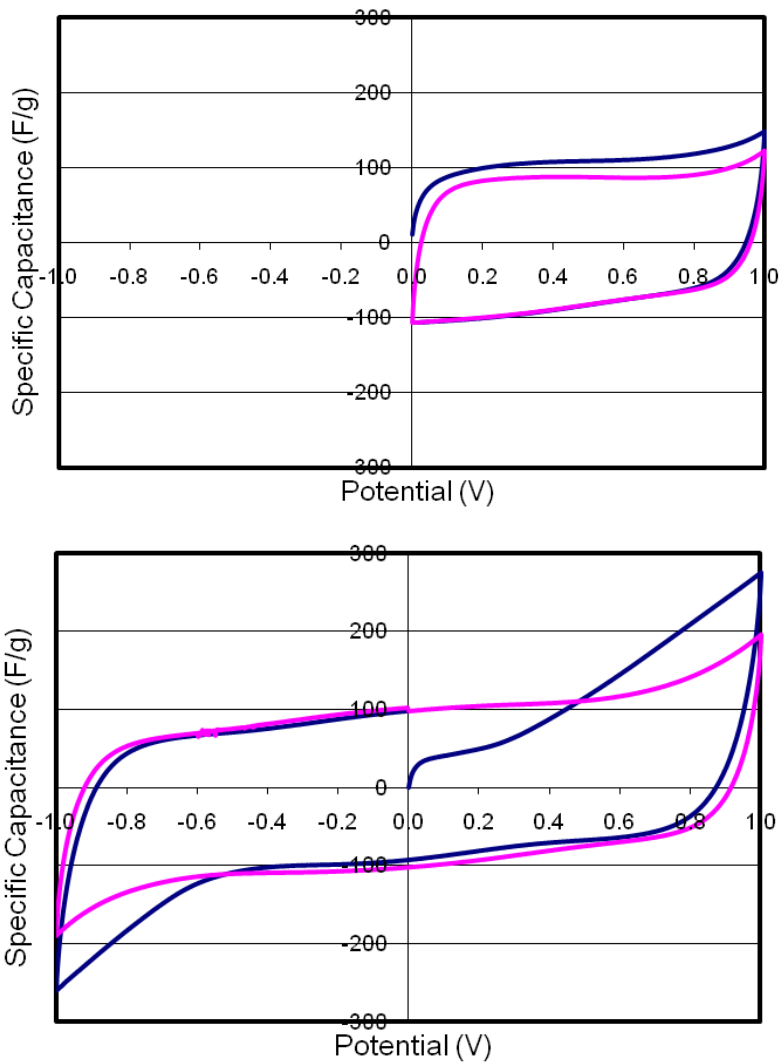

Figure 6. CV from 0.0 V to 1.0 V (top) and from -1.0 V to 1.0 V (bottom)

Table 4. Comparison of methods to calculate specific capacitance from CV curves.

| Method that current was determined | Csp |
|---|---|
| Maximum current $1^{st}$ scan (CV from -1 to 1V) | 275.2 F/g |
| Maximum current $20^{th}$ scan (CV from -1 to 1V) | 155.6 F/g |
| Maximum current $1^{st}$ scan (CV from 0 to 1V) | 152.1 F/g |
| Current at zero V $20^{th}$ scan (CV from -1 to 1V) | 105.0 F/g |
| Ave discharge current $20^{th}$ scan (CV from -1 to 1V) | 102.6 F/g |
| Ave discharge current $20^{th}$ scan (CV from 0 to 1V) | 77.3 F/g |